\documentclass[11pt]{paper}

\usepackage[english]{babel}

\usepackage{amsmath}
\usepackage{graphicx}
\usepackage[table]{xcolor}
\usepackage[top=1in, bottom=1in, left=1in, right=1in]{geometry}

\usepackage[table]{xcolor}
\usepackage[para]{footmisc}
\usepackage{lmodern}
\usepackage[T1]{fontenc}
\usepackage{graphicx}
\usepackage{rotating}
\usepackage[utf8]{inputenc}
\usepackage[english]{babel}
\usepackage{fancyhdr}
\usepackage{bm}
\usepackage{wrapfig}
\usepackage{pdfpages}
\usepackage{array}
\usepackage{setspace}
\usepackage{appendix}
\usepackage{tikz}

\usepackage{amssymb}
\usepackage{amsfonts}
\usepackage{fancybox}
\usepackage{fancyhdr}
\usepackage{yfonts}
\usepackage{wrapfig}
\usepackage{enumitem}
\usepackage{xpatch}

\usepackage[colorlinks = true,
            linkcolor = blue,
            urlcolor  = blue,
            citecolor = blue,
            anchorcolor = blue]{hyperref}

\makeatletter
\newcommand{\srcsize}{\@setfontsize{\srcsize}{5pt}{5pt}}
\makeatother

\begin{document}





\title{Early Career Issues in Particle Physics}
\author{Saptaparna Bhattacharya, Southern Methodist University \\ 
\small{Presented at the 32nd International Symposium on Lepton Photon Interactions at High Energies,}\\
\small{Madison, Wisconsin, USA, August 25-29, 2025}}

\maketitle

\section*{\textit{Disclaimer}}

{\textit{These proceedings are based on a plenary presentation that I gave at the \href{https://indico.cern.ch/event/1493037/timetable/?view=standard}{Lepton Photon conference} in August, 2025. All opinions shared in this document are mine alone. I selected a vantage point with which I have utmost familiarity, therefore, the talk was given from the perspective of a collider physicist. Consequently, some aspects of my talk are specific to collider physics, while others transcend particular research areas (such as, physics goals, energy regimes, and research facilities) and are applicable to particle physics in general.}}

\section{Introduction}

Particle physics stands at a crossroads. The discovery of the Higgs boson in 2012 completed the Standard Model, but it also opened a new set of fundamental questions about the nature of our universe. Answering these questions, from the hierarchy problem and the nature of dark matter to the origin of neutrino masses, will require a new generation of powerful and sophisticated experiments. The global community is currently engaged in a comprehensive, long-term planning process to determine the ``next big machine''.

These discussions are not merely technical. The decisions made in the coming years will shape the field until 2060 and beyond. This extended timeline has a direct and immediate impact on the lives and careers of the field's youngest members: the graduate students, postdoctoral fellows, and other early-career researchers (ECRs) who will be tasked with designing, building, and ultimately analyzing the data from these future facilities.

This proceeding provides an overview of the key issues facing ECRs in particle physics today, based on a talk delivered on the subject. It begins by surveying the future collider landscape from an ECR perspective. It then delves into the critical feedback gathered directly from ECRs through large-scale surveys, focusing on their primary concerns and actionable recommendations. Following this, it presents an analysis of the academic job market, the important but often unrewarded role of science outreach, and the emerging dual-edged challenge and opportunity of artificial intelligence. The paper concludes by highlighting the indispensable support networks available to ECRs, which are vital for navigating this complex and uncertain career path.

\begin{figure}[!htb]
\centering
\includegraphics[width=15.0cm]{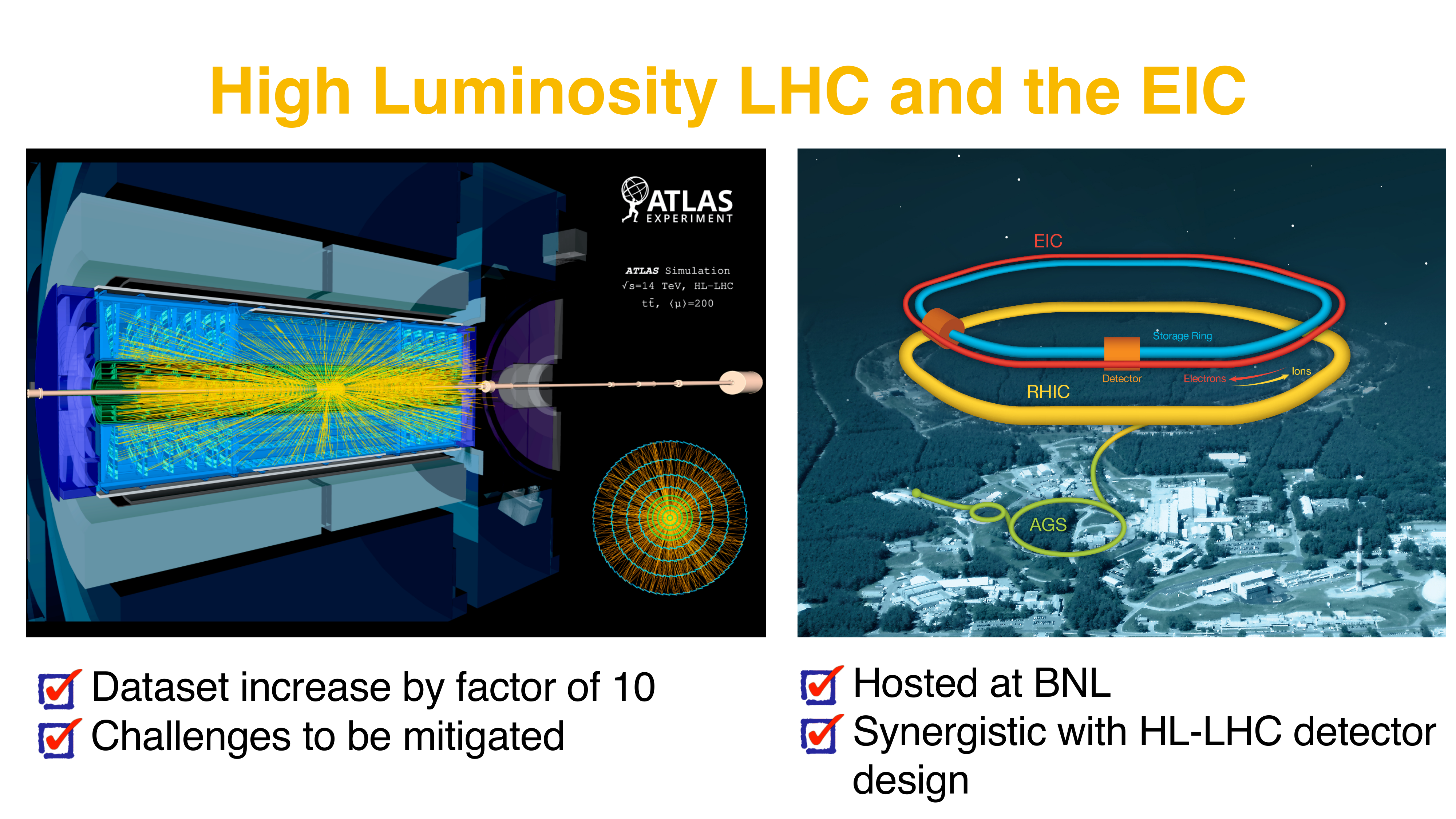}
\caption{The High-Luminosity LHC and the Electron Ion Collider are the next big projects on the horizon}
\label{fig:certainty}
\end{figure} 

\section{The Future Collider Landscape}

From an ECR's perspective, the future of collider physics is a mixture of concrete certainty and ambitious, long-term possibilities. In the immediate future, the field is anchored by two major funded projects. The \textbf{High-Luminosity LHC (HL-LHC)} at CERN is the highest priority, promising a tenfold increase in the LHC dataset. This upgrade presents immense challenges in data handling, simulation, and analysis, providing a rich research program for the next two decades. In the United States, the \textbf{Electron-Ion Collider (EIC)}, to be built at Brookhaven National Laboratory, will open a new frontier in nuclear physics, exploring the structure of protons and nuclei with unprecedented precision. Its detector R\&D and design have synergistic connections with the HL-LHC, providing concrete opportunities for ECRs as shown in Fig.~\ref{fig:certainty}.

\begin{figure}[!htb]
\centering
\includegraphics[width=15.0cm]{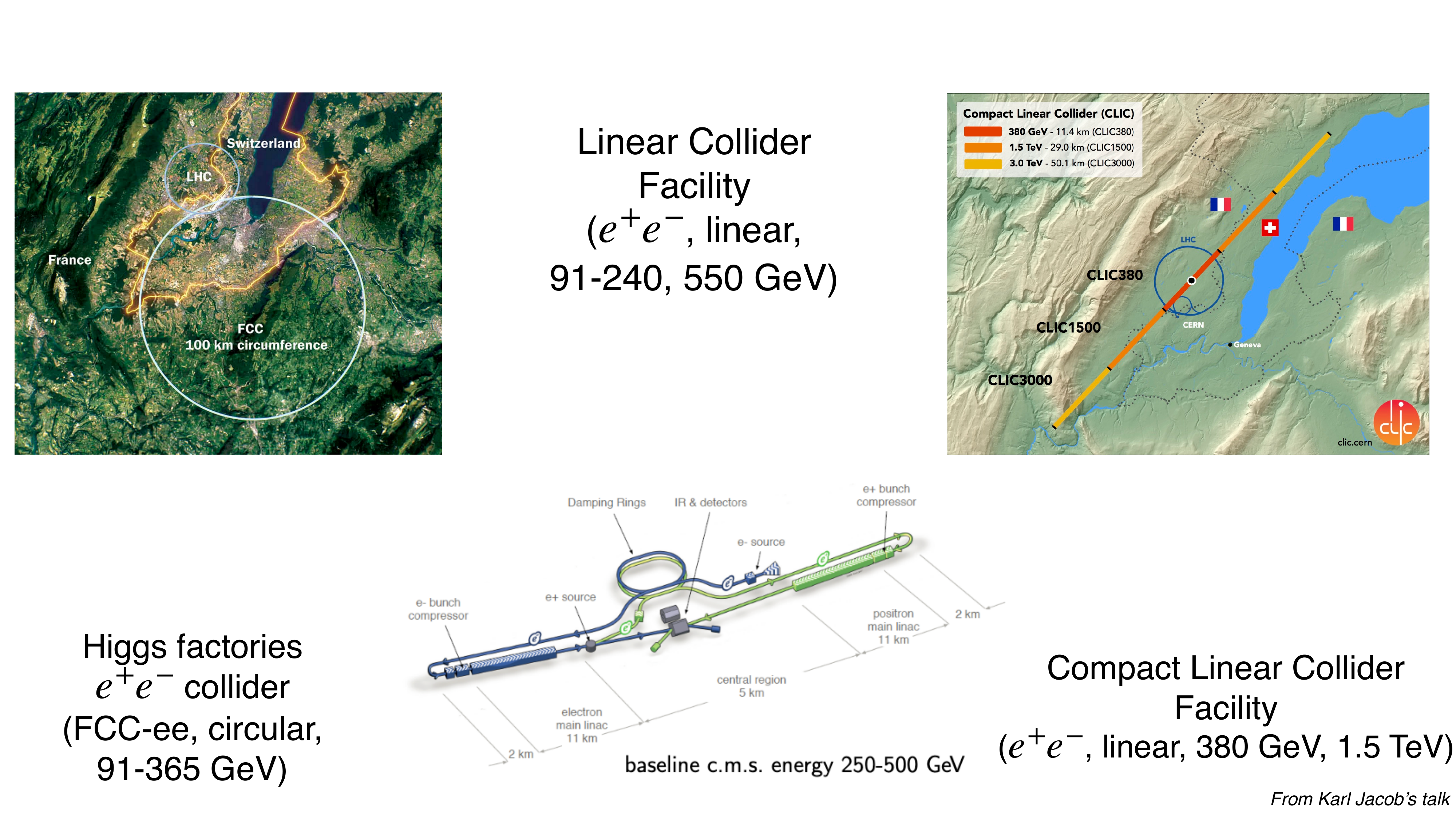}
\caption{Various future scenarios}
\label{fig:future}
\end{figure} 

Beyond these certainties lies a map of ambitious proposals for the post-LHC era as shown in Fig.~\ref{fig:future}, often broadly categorized as ``Higgs factories'' or ``energy frontier'' machines.

\begin{itemize}
    \item \textbf{Circular $e^+e^-$ Colliders:} The leading proposal is the \textbf{Future Circular Collider (FCC-ee)} at CERN, envisioned as a 100 km circular tunnel that would first host an electron-positron collider for precision Higgs and electroweak measurements, before potentially housing a 100 TeV proton-proton collider (FCC-hh) in the same tunnel.

    \item \textbf{Linear $e^+e^-$ Colliders:} Both the \textbf{International Linear Collider (ILC)}, proposed to be hosted in Japan, and the \textbf{Compact Linear Collider (CLIC)} at CERN offer pathways to a high-energy, precision $e^+e^-$ environment. These linear designs provide a different set of technical challenges and opportunities compared to their circular counterparts.

    \item \textbf{The Muon Collider:} A Muon Collider has emerged as a particularly exciting high-reward option. By colliding muons, which are 200 times heavier than electrons, a muon collider could reach collision energies of 10 TeV or more in a tunnel significantly smaller than the FCC. This machine offers a direct path to the high-energy frontier and precision Higgs physics, such as measuring the Higgs potential. The recent 2025 US National Academies report strongly endorsed an aggressive R\&D program to prove the feasibility of this concept.
\end{itemize}

\begin{figure}[!htb]
\centering
\includegraphics[width=15.0cm]{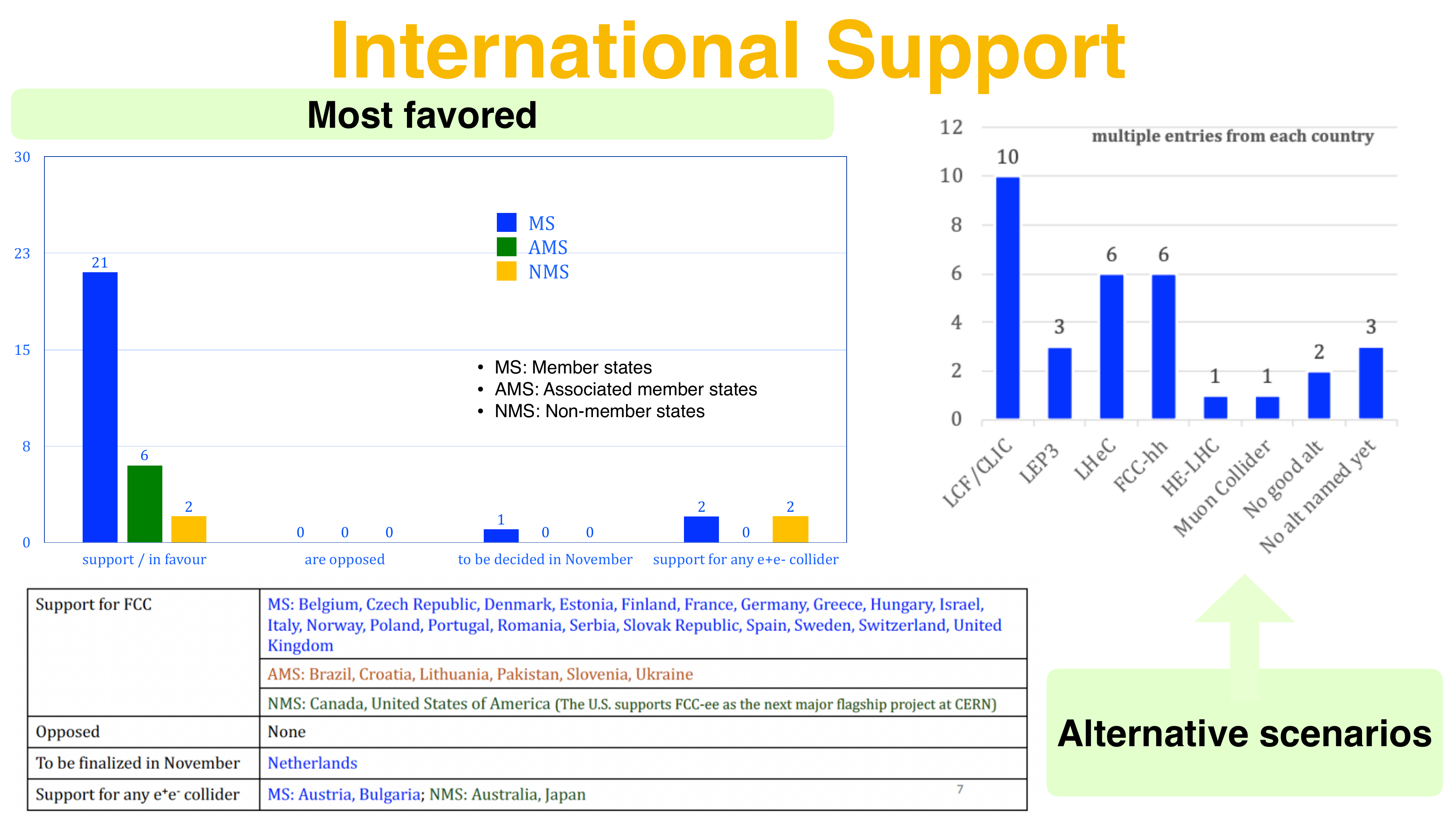}
\caption{International support for the next collider}
\label{fig:support}
\end{figure}

The community's support for these various options is still coalescing, as seen in inputs to the 2025 European Strategy for Particle Physics (ESPP) update. While the Muon Collider generates significant excitement (garnering a 4.6/5 mean response in one survey~\cite{USECR}), the FCC-ee and linear colliders are seen as intermediate viable paths forward providing unparallel opportunities for precision physics. The decisions made by bodies like the CERN Council, informed by the ESPP, will lock in a direction for generations. There is ample international support for a ``Higgs-factory" can seen in Fig.~\ref{fig:support}.

\section{The Importance of Community-Driven Planning}

Given the immense cost and decades-long timelines of these projects, community consensus is essential. The particle physics community has established a ``bottom-up''  approach to strategic planning through decadal surveys.

In the United States, the ``Snowmass''  process, organized by the APS Division of Particles \& Fields (DPF), brings together thousands of physicists to identify scientific priorities and opportunities across multiple frontiers (Energy, Neutrino, Cosmic, etc.). The 2022 Snowmass process produced a consensus set of recommendations, which were then used by the High Energy Physics Advisory Panel (HEPAP) to create the P5 (Particle Physics Project Prioritization Panel) report. The 2023 P5 report's recommendations included:

\begin{enumerate}
    \item Full support for the HL-LHC as the highest-priority large project.
    \item Participation in an offshore $e^+e^-$ Higgs factory in collaboration with international partners.
    \item An R\&D program for a 10 TeV Muon Collider
    \item Strengthening computational physics and precision theory calculations.
\end{enumerate}

While the P5 report's recommendations are far reaching, the above list is specific to collider physics. 

\begin{figure}[!htb]
\centering
\includegraphics[width=15.0cm]{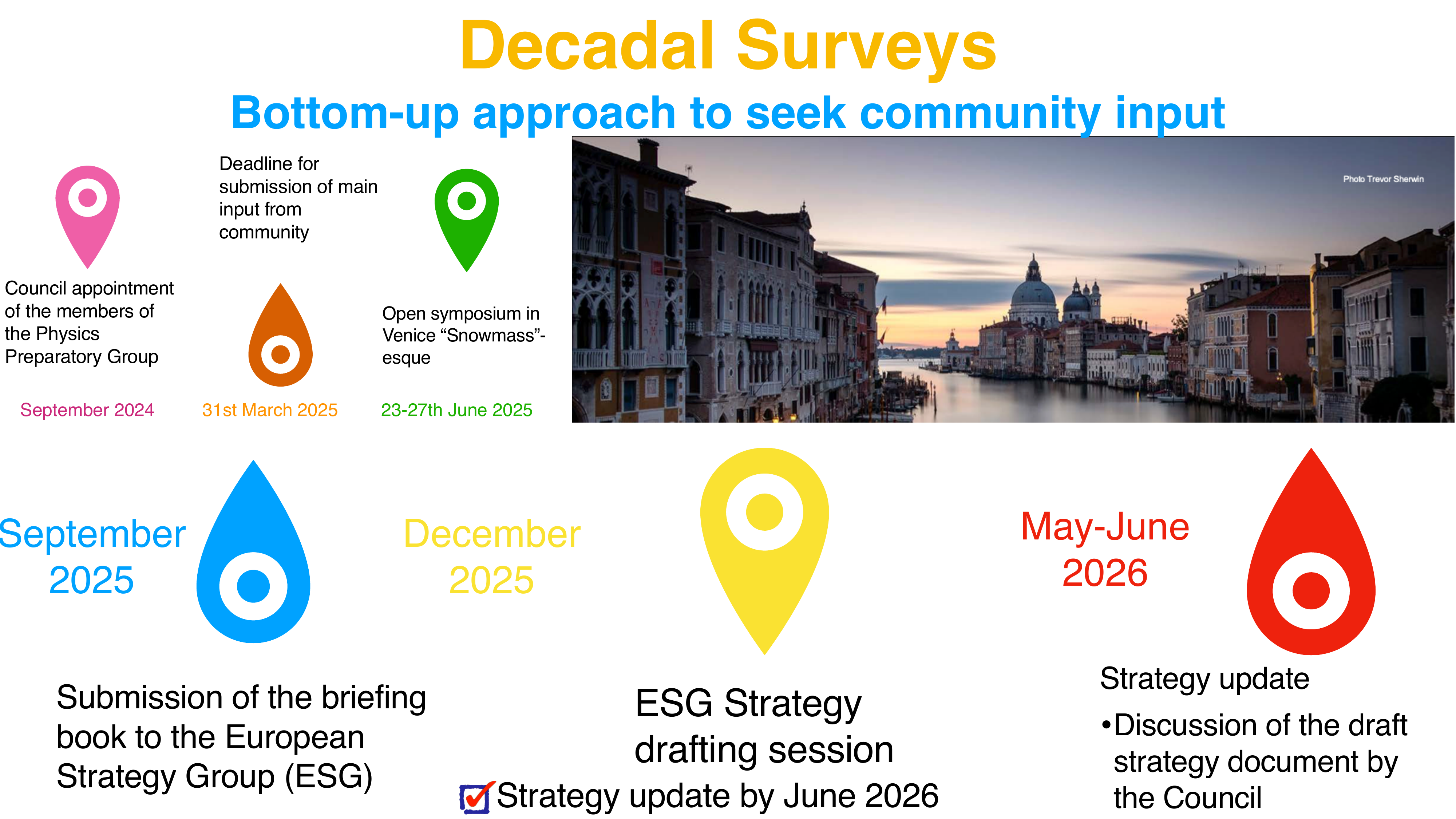}
\caption{The community driven European Strategy Update process}
\label{fig:esppu}
\end{figure}

This bottom-up, community-driven process is not unique to the US. The European Strategy for Particle Physics (ESPP) follows a similar model (as shown in Fig.~\ref{fig:esppu}), with an open symposium and community-submitted white papers informing the European Strategy Group (ESG). The next ESPP update is scheduled for release by June 2026.

The impact of these organized, clear-eyed reports cannot be overstated. The 2014 P5 report, for example, was widely praised for its clarity and focus. It set a clear direction for the US HEP program, which led to stable funding and the successful launch of new projects like LBNF/DUNE. For ECRs, this kind of strategic clarity is vital. It provides a roadmap that allows them to align their research interests and skill development with the future direction of the field, offering a measure of stability in an otherwise uncertain landscape.

\section{Early Career Researcher Voices: Key Concerns}

As part of the 2025 ESPP update, ECRs organized to produce a dedicated white paper~\cite{ECR} to ensure their voices were heard in the planning process. This effort, grounded in a community-wide survey of over 800 ECRs across Europe, provides a stark and quantitative look at the most pressing issues facing the next generation.

The survey demographics revealed that the largest respondent groups were PhD students (42.1\%) and postdoctoral researchers (31.9\%). While a majority (81\%) reported feeling heard within their local research groups, this number dropped to just 54\% when asked about their large, international collaborations (e.g., ATLAS, CMS), indicating a disconnect between local environment and the larger institutional structures that govern the field.

The most critical issues identified by ECRs relate directly to career stability and mental well-being:

\begin{itemize}
    \item \textbf{Lack of Long-Term Stability:} The primary concern is the general lack of long-term planning and stability, which is exacerbated by the long timelines of future projects and the precarious nature of short-term postdoc contracts.
    \item \textbf{Scarcity of Permanent Positions:} A significant majority of respondents wish to remain in academia, but only 45\% consider their chances of securing a permanent position to be ``okay or pretty good.''  This gap between aspiration and perceived reality is a major source of stress.
    \item \textbf{Mental Health:} The survey confirmed that the mental health of ECRs is a topic of serious concern, with rates of depression and anxiety among this population consistently exceeding those of the general public. The most significant measures identified to improve personal well-being were, unsurprisingly, more job security and greater location stability.
\end{itemize}

These findings paint a clear picture: ECRs are passionate about their research but are struggling under a system that demands geographic mobility and offers a very narrow and uncertain path to a stable, permanent career.

\section{Early Career Researcher Voices: Recommendations}

The ECR white paper did not just identify problems; it formulated 55 actionable recommendations for the community, institutions, and senior researchers to implement. These recommendations aim to improve the future of ECRs and, by extension, the future of the field itself. Two key recommendations stand out:

\begin{enumerate}
    \item \textbf{Increase Awareness of Employment Perspectives:} The academic community, particularly senior researchers, PIs, and institutions, must take active steps to increase awareness about the realities of the academic job market. This transparency should begin at the earliest stages of an ECR's career (i.e., during their PhD). Students should be given a realistic understanding of their career prospects both inside and outside academia, rather than being implicitly guided only toward a tenure-track path that few will attain.
    \item \textbf{Integrate Transferable Skills into Training:} ECR training programs, especially PhD projects, should be restructured to formally integrate skills that are transferable to industry. Supervisors should be required to explicitly outline these skills (e.g., advanced data analysis, machine learning, project management, complex instrumentation) in their project proposals. This change would serve a dual purpose: it would make ECRs more competitive for jobs in industry, and it would formally validate the non-academic career path as a successful outcome of a physics PhD.
\end{enumerate}

These recommendations call for a cultural shift—one that moves away from the ``tenure-track or bust''  mentality and toward a more holistic view of career development that equips ECRs for success, wherever their path may lead.

\section{Navigating the Academic Job Market}

An analysis of the High Energy Physics (HEP) academic job market, using data from the community-run ``rumor mill,''  confirms the precarity felt by ECRs. The data reveals several key trends:

\begin{itemize}
    \item \textbf{High Volatility:} The total number of filled faculty positions varies significantly from year to year. For example, after a peak in 2016, the market saw a sharp $\sim$26\% drop in advertised positions during the 2020-2021 cycle, likely due to the COVID-19 pandemic, before recovering in subsequent years.
    \item \textbf{Trend Toward Generalization:} In recent years, there has been a noticeable trend away from jobs tied to a specific experiment (e.g., ``LHC``) and toward more generalist ``data analysis''  or ``computational frontier''  positions.
    \item \textbf{Rise of AI/ML:} A proliferation of jobs specifically advertising for ``AI/ML''  expertise has been observed in the 2024-2025 cycle, reflecting a broader shift in the field's priorities.
\end{itemize}

\subsection{Analysis of the job market based on harvesting data}

\begin{figure}[!htb]
\centering
\includegraphics[height=7.0cm]{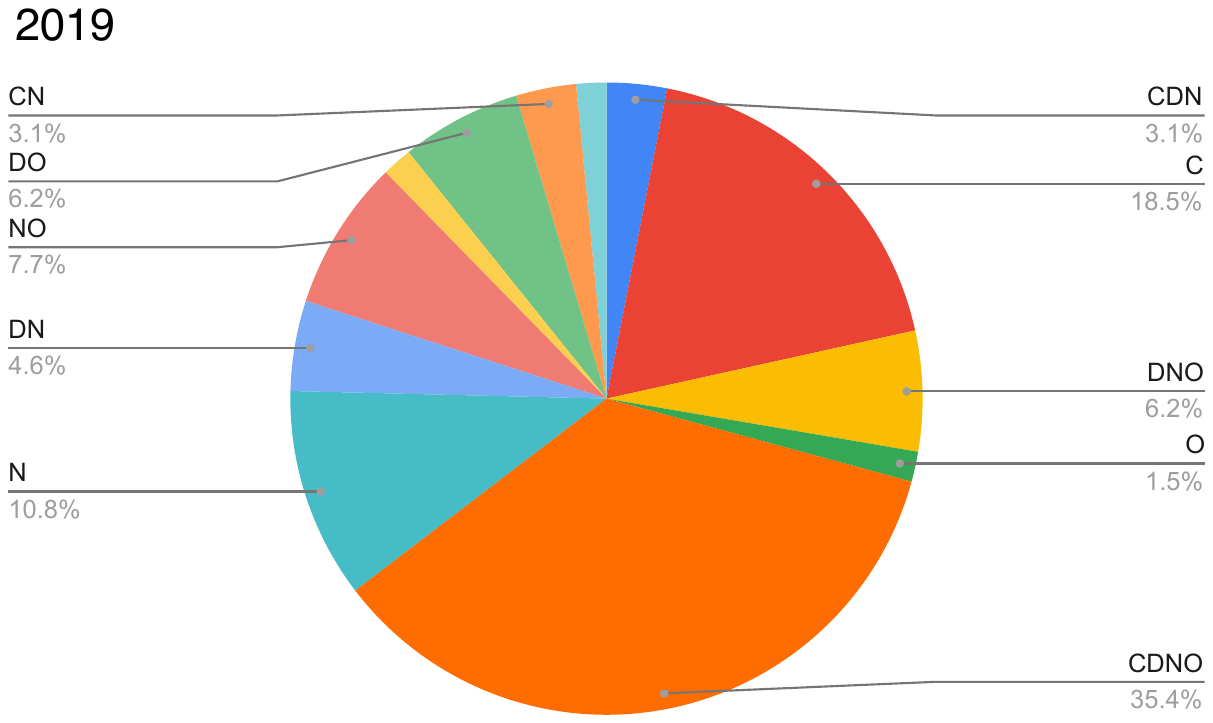}
\includegraphics[height=7.0cm]{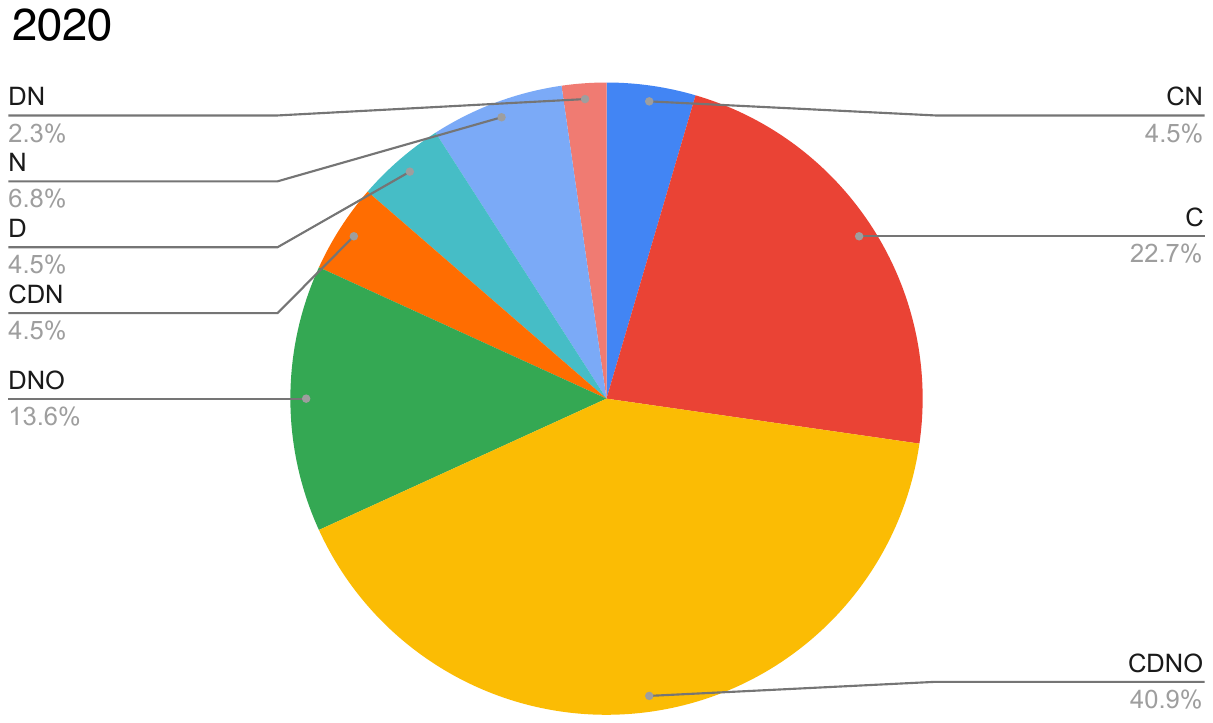}
\includegraphics[height=7.0cm]{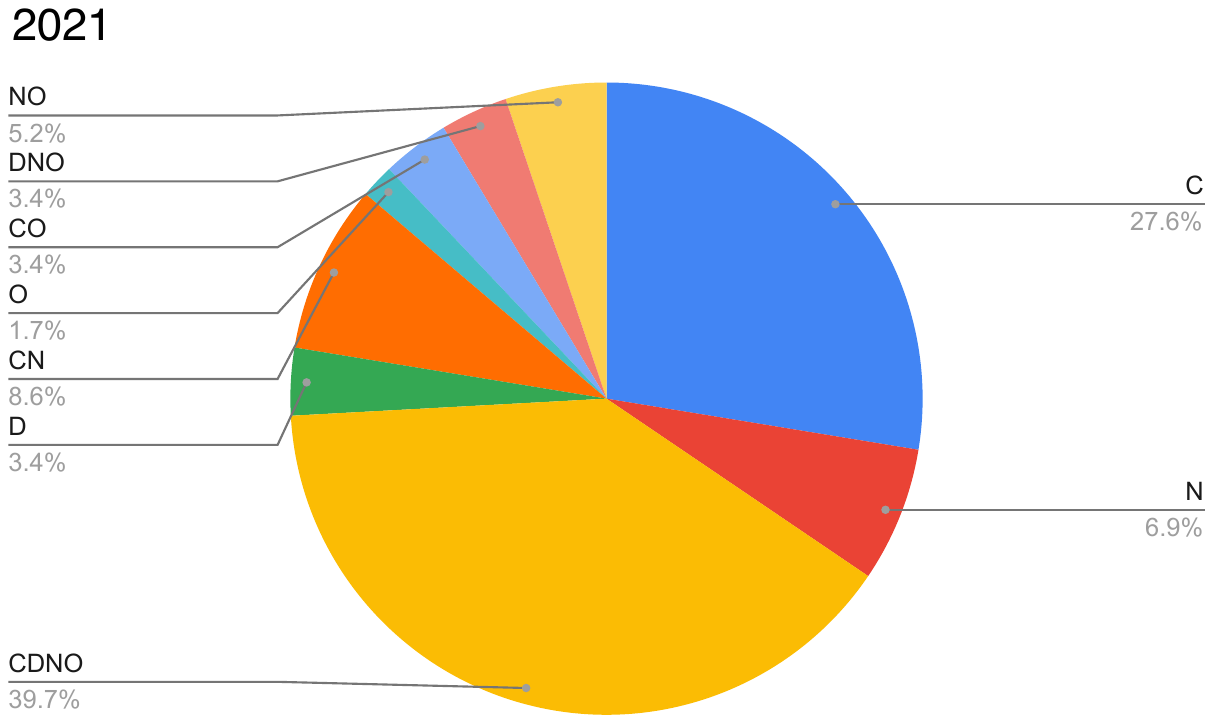}
\caption{Job market variations: 2019-2021}
\label{fig:job_market1}
\vspace{-0.3in}
\end{figure}

\begin{figure}[!htb]
\centering
\includegraphics[height=7.0cm]{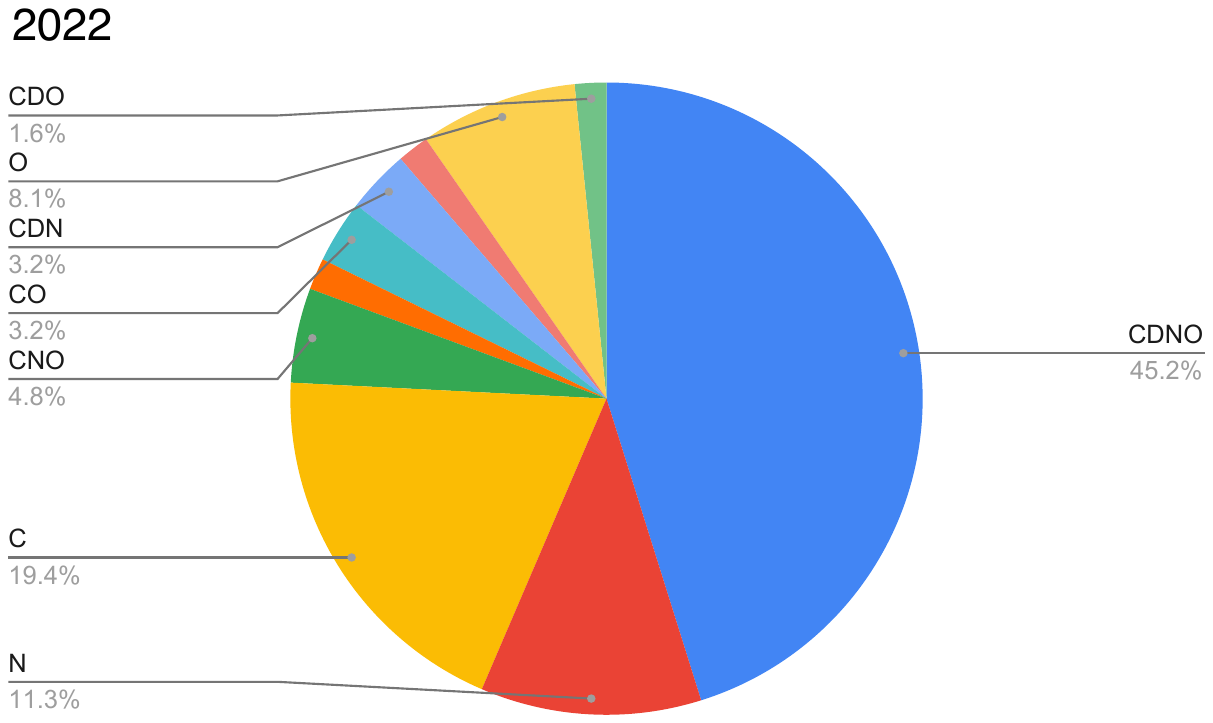}
\includegraphics[height=7.0cm]{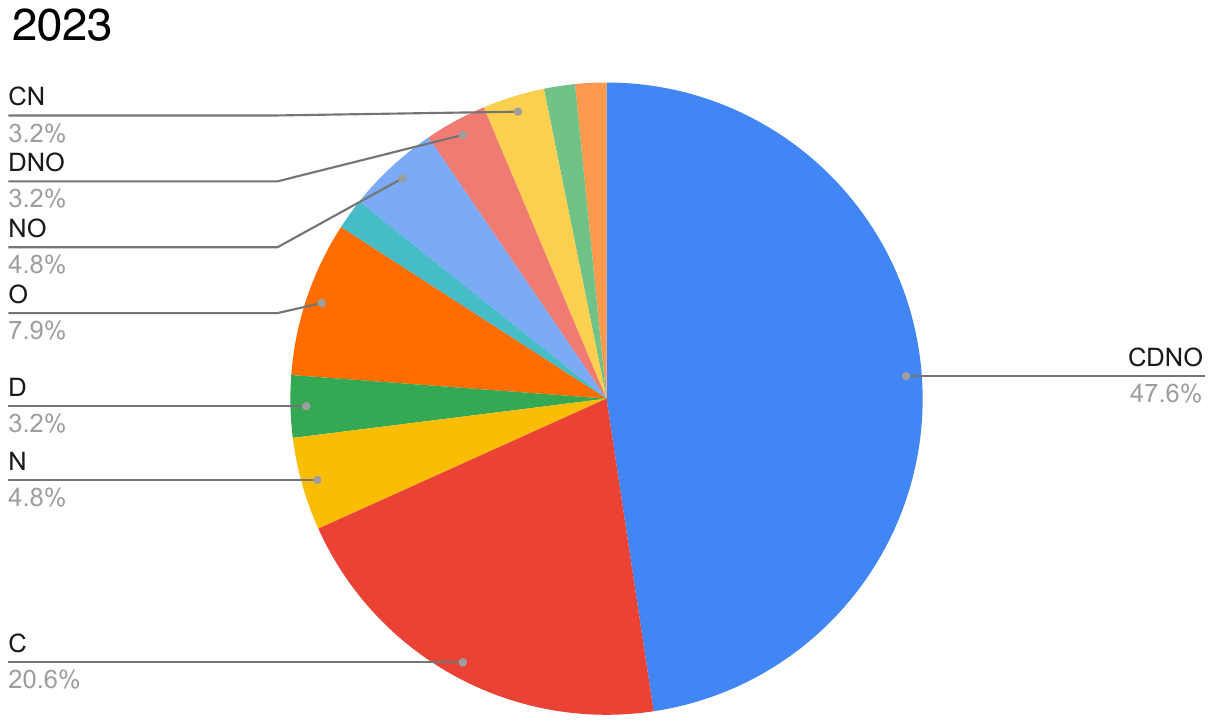}
\includegraphics[height=7.0cm]{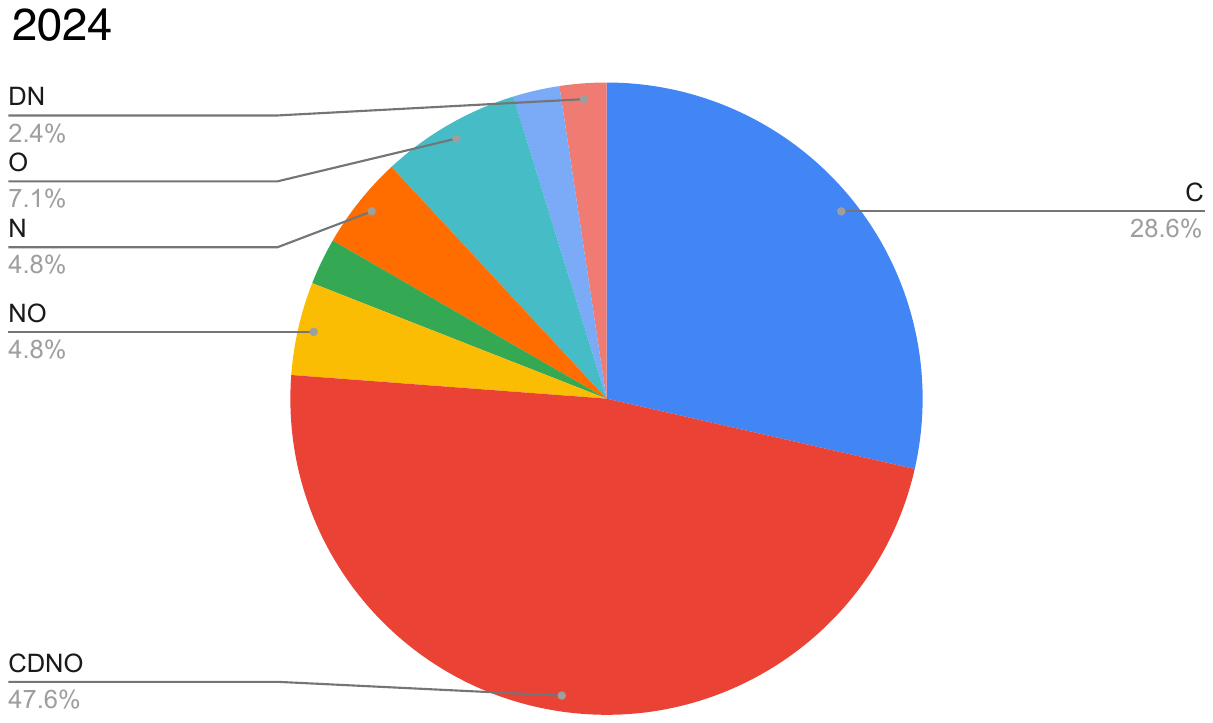}
\caption{Job market variations: 2022-2024}
\label{fig:job_market2}
\vspace{-0.2in}
\end{figure}

The following pie-charts shown in Fig.~\ref{fig:job_market1} and \ref{fig:job_market2} show year-wise variation in field-specific job advertisements. The following short-hand is used to designate each field: ``C'': Collider-specific job, ``D'': Dark-matter or relatively smaller experiment centric job, ``N'': job with a focus on neutrino physics and finally, ``O'': catch-all label for all other possibilities, including jobs that focus on AI/ML. Jobs are often advertised with multiple labels as shown in Fig.~\ref{fig:job_market1}, \ref{fig:job_market2}.

The data for theory faculty hires shows a similar volatility and, notably, a long ``time-to-faculty''  of 5-6 years post-PhD on average. My personal career journey from a PhD at Brown University, through postdoctoral positions at Northwestern and DESY, a ``pathway-to-faculty''  fellowship at Wayne State, and finally to an Assistant Professor position at SMU, serves as a case study. This multi-step, geographically diverse path spanning nearly a decade post-PhD is typical and highlights the persistence and mobility required to navigate the academic ladder.

\section{The Role of Outreach: A Disconnect in Recognition}

Public outreach and science communication are critical for the long-term health of particle physics, especially as the field advocates for publicly-funded, multi-billion-dollar future experiments. ECRs are exceptionally active in this area.

\begin{figure}[htbp]
\centering
\includegraphics[width=7.0cm]{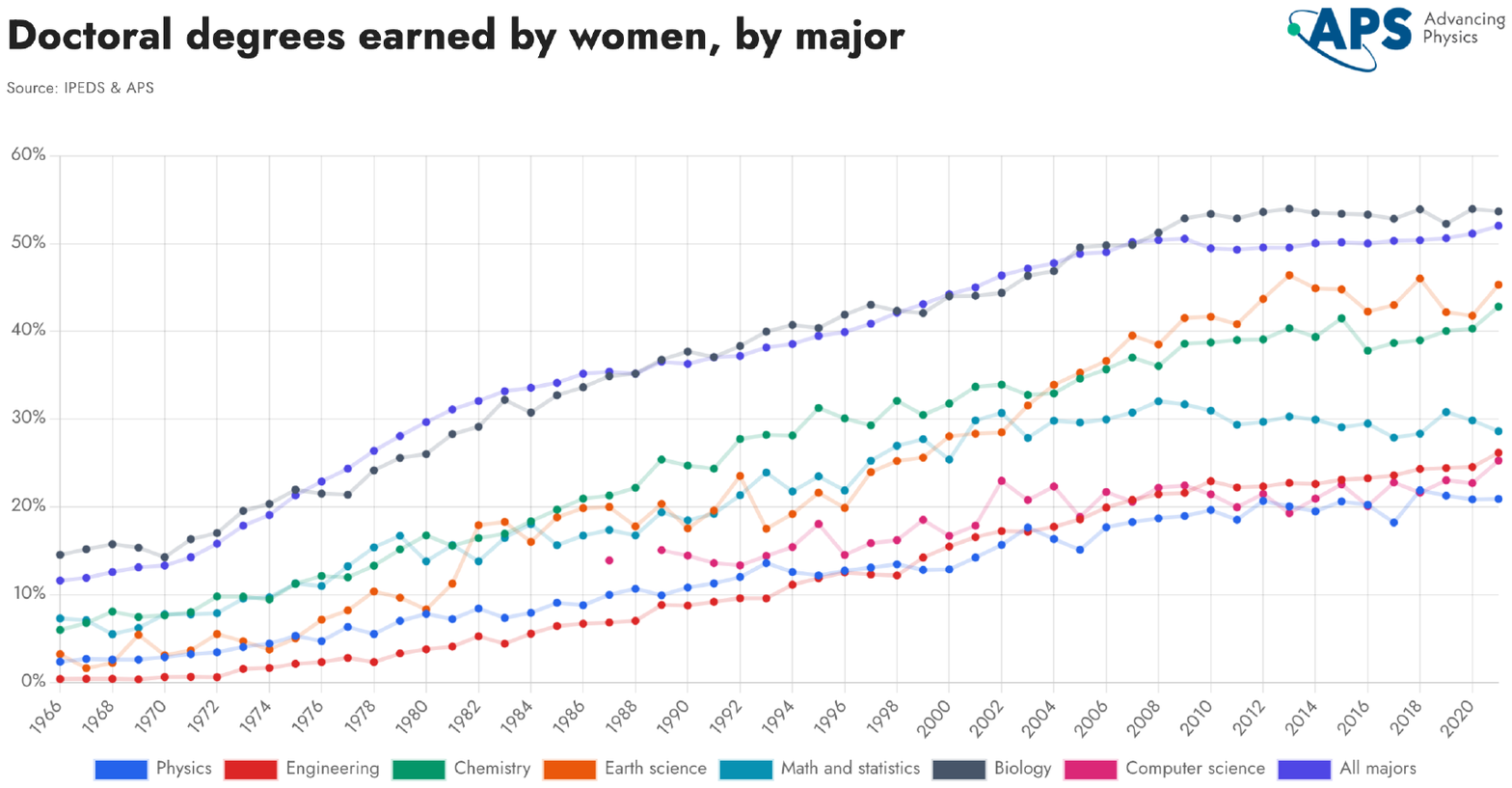}
\includegraphics[width=8.0cm]{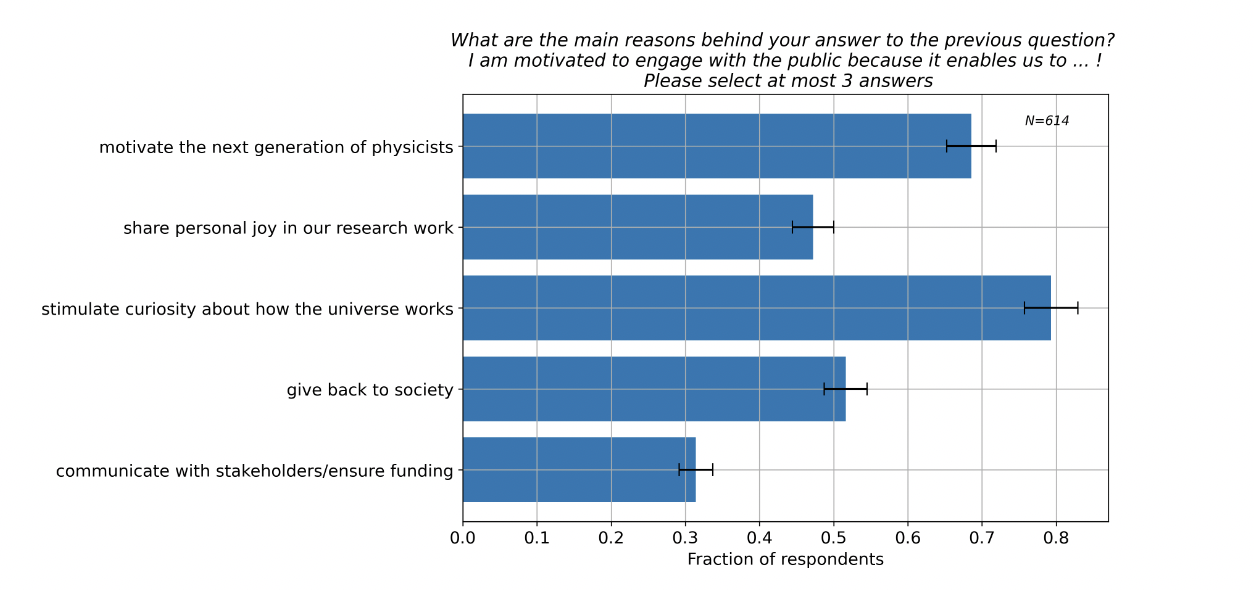}
\caption{Left: Women in physics (data from the \href{https://www.aps.org/learning-center/statistics/education}{APS website}), Right: Motivation for outreach based on survey in Ref.~\cite{ECR}}
\label{fig:aps}
\end{figure} 

A survey presented in the ECR white paper found that a significant majority (84\%) of ECRs are motivated to engage with the public (as shown on the right panel of Fig.~\ref{fig:aps}). Their primary motivations are not to ``ensure funding''  (the lowest-ranked reason), but rather to ``stimulate curiosity about how the universe works''  (the top reason), ``motivate the next generation of physicists,''  and ``share personal joy in our research work''.

However, a significant disconnect exists between the value of this work and its formal recognition. This is particularly stark when viewed through the lens of diversity.

\begin{itemize}
    \item \textbf{The Gender Disparity:} While women earn approximately 20\% of doctoral degrees in physics (as shown on the left panel of Fig.~\ref{fig:aps})—a number that has stubbornly plateaued for two decades—their participation in outreach and education efforts is disproportionately high, estimated at up to 70\% in some contexts.
    \item \textbf{Lack of Formal Reward:} This ``service''  work, which is essential for the field's public image and talent pipeline, is rarely, if ever, a significant factor in performance evaluations, funding applications, or job searches.
\end{itemize}

This leads to a key ECR recommendation: \textbf{Systemic changes should be made to formally recognize and reward outreach and communication efforts.} These activities must be integrated into institutional benchmarks, performance reviews, and funding evaluations to validate their importance within the scientific profession. Until this happens, outreach will remain a passion project that ECRs—particularly women and underrepresented minorities—are expected to do for free, on top of their already demanding research and teaching responsibilities.

\section{New Challenges and Opportunities: Artificial Intelligence}

The rapid ascent of generative Artificial Intelligence (AI) presents both a novel challenge and an enormous opportunity for the particle physics community.

\subsection*{The Challenge: Teaching in the Age of AI}
As newly minted educators, ECRs are on the front lines of a rapidly changing pedagogical landscape. Generative AI tools like Gemini or ChatGPT can now produce complex, detailed, and largely correct physics analysis plans from a simple prompt. An example request to ``Construct an analysis separating the WH and WZ process''  can yield a multi-page document detailing key variables ($m_{bb}$, $E_T^{\text{miss}}$), background processes, and even the formula for statistical significance.

This new technology challenges traditional assessment methods. As a ``coping mechanism,''  educators must adapt. One proposed policy is to permit the use of generative AI, but to require students to cite its use explicitly, including the exact prompts used and a description of the AI's output. This promotes transparency and treats AI as a tool to be used critically, rather than a forbidden source of plagiarism.

\begin{figure}[!htb]
\centering
\includegraphics[width=15.0cm]{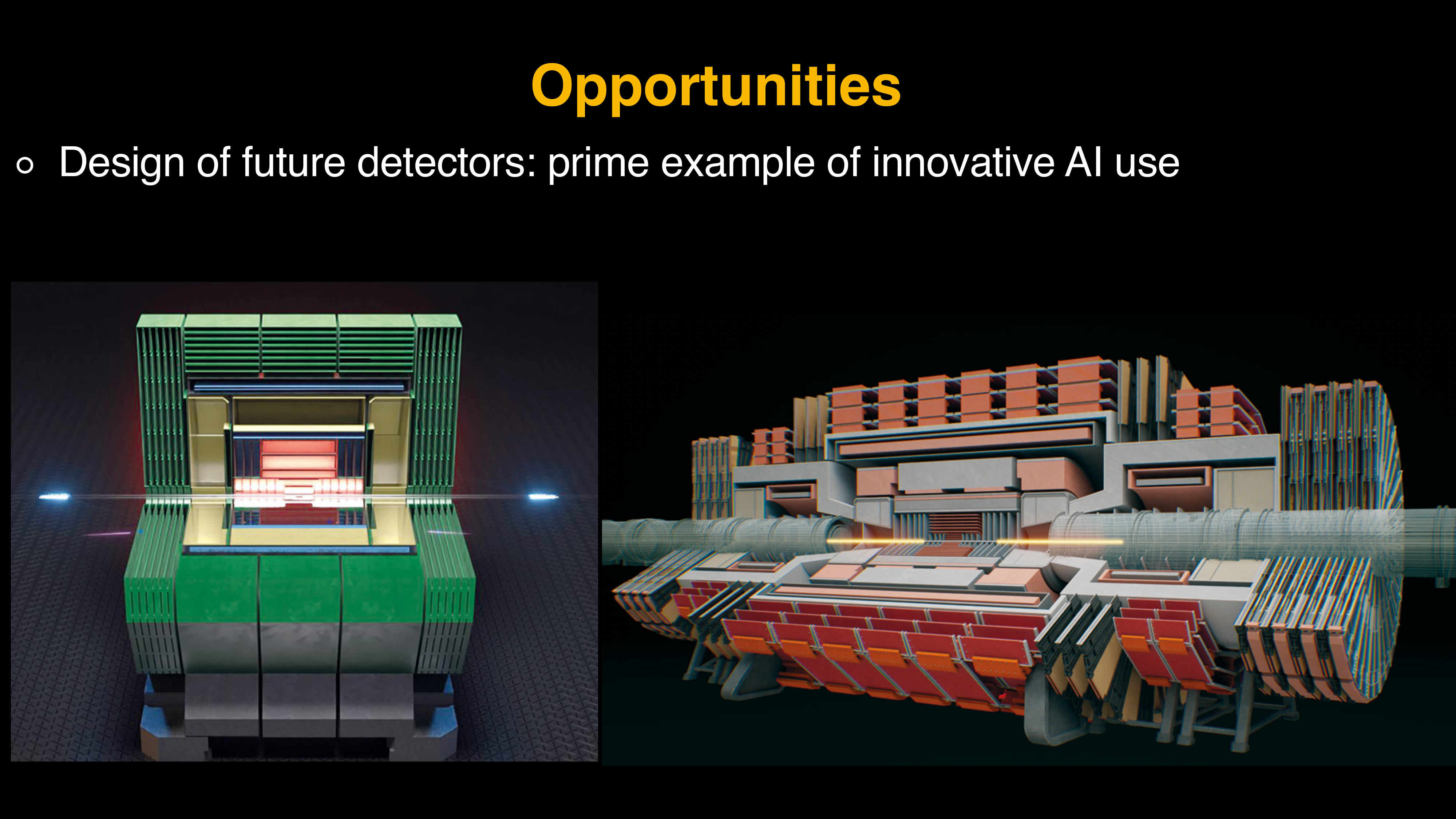}
\caption{Detectors of the future}
\label{fig:det}
\end{figure} 

\subsection*{The Opportunity: AI for Science}
Conversely, AI and Machine Learning (ML) represent one of the greatest opportunities for the field. The US government has recognized this, with legislation proposing \$150 million for the DOE to develop ``self-improving artificial intelligence models for science and engineering''. This initiative includes the creation of an ``American science cloud''  to provide these tools to the research community.

For particle physics, the applications are boundless:
\begin{itemize}
    \item \textbf{Data Analysis:} AI/ML techniques are already at the heart of event reconstruction, simulation, and data analysis at the LHC.
    \item \textbf{Detector Development:} AI can be used to optimize detector geometry, creating a positive feedback loop that studies the impact of sensor placement and material on physics performance, thereby improving the design itself as shown in Fig.~\ref{fig:det}.
    \item \textbf{Career Paths:} This boom in AI/ML creates new, highly-valued career paths for ECRs, both within academia (as seen in recent job postings) and in industry.
\end{itemize}

\section{Support Structures for Early Career Researchers}

Given this complex landscape of long-term uncertainty and high personal stakes, it is essential for ECRs to know that they are not alone. A robust ecosystem of professional societies and user organizations exists to provide support, community, and advocacy.

\begin{itemize}
    \item \textbf{APS Division of Particles \& Fields (DPF):} The DPF is an excellent way for ECRs to connect with the broader US particle physics community. It provides opportunities to serve on committees, engage in HEP advocacy and funding efforts, and participate in mentorship programs. These include the Industry Mentoring for Physicists (IMPact) program, which connects ECRs with industrial physicists, and the National Mentoring Committee, which provides a support network for students.
    \item \textbf{Laboratory User Organizations:} Groups like the SLAC Users Organization (SLUO), the Fermilab Users Executive Committee (UAEC), and the US LHC Users Association (US LUA) are crucial support systems. They provide direct support to ECRs in the form of:
    \begin{itemize}
        \item Mentorship programs.
        \item Conference and travel support.
        \item Advocacy: Sponsoring ``lightning round''  speakers to travel to Washington D.C. to advocate for sustained science funding.
        \item Logistical Support: Helping researchers navigate the practicalities of life at international labs like CERN.
    \end{itemize}
\end{itemize}

These organizations empower ECRs, give them a collective voice, and provide the community and resources needed to sustain a challenging career.

\section{Conclusion}

The field of particle physics is driven by fundamental questions that span generations. The pursuit of answers requires a multi-decade vision for future experiments, a vision that is currently being forged by the global community. Early-career researchers are the most vital component of this vision, as they will be the ones to bring it to fruition.

As this paper has shown, ECRs are deeply passionate and motivated, but they also face significant and well-documented stressors related to budgetary constraints, project delays, and profound job uncertainty. Their concerns, articulated through community-wide surveys, are not just complaints; they are a clear-eyed diagnosis of the field's sustainability problem.

The path forward must be one of partnership. The community must heed the recommendations of its ECRs: to provide greater transparency in career paths, to formally value and integrate transferable skills, and to systemically recognize and reward essential work like outreach. Simultaneously, ECRs must seize the many opportunities available to them in detector R\&D, software, and the burgeoning field of AI, while leaning on the robust support networks provided by the APS and laboratory user organizations.

By acknowledging these challenges head-on and committing to a culture that actively supports its youngest members, the particle physics community can ensure that it remains a vibrant, innovative, and sustainable field for generations to come.

\end{document}